\documentstyle[12pt]{article}

\title{Fractional Statistics in Three Dimensions: Compact
Maxwell-Higgs System}
\author{H.~Fort and R.~Gambini  \\
{\sl
Instituto de F\'{\i}sica, Facultad de Ciencias, Universidad de la}\\
{\sl Rep\'ublica. Trist\'an Narvaja 1674, 11200 Montevideo, Uruguay.}\\
}

\date{\today}

\begin{document}
\bibliographystyle{bibstand}
\maketitle

\begin{abstract}

We show that a (3+1)-dimensional system composed of an open
magnetic vortex and an electrical point
charge exhibits the phenomenon of Fermi-Bose transmutation.
In order to provide the physical realization of this system
we focus on the
lattice compact scalar electrodynamics $SQED_c$ whose topological
excitations are open Nielsen-Olesen strings with magnetic monopoles
attached at their ends.
\end{abstract}


\newpage

                Some years ago it was showed that
a system composed of a Nambu charged closed string interacting
with a Kalb-Ramond \cite{kr} point source
display fractional statistics \cite{gs}.
In a previous paper \cite{fg}
we addressed the issue of supplying the physical realization
of this, in principle, {\em ad hoc} system.
We found that a closed Nielsen-Olesen ({\em N-O}) string
vortex \cite{no} in presence of an ordinary electric charge
exhibits the phenomenon of Fermi-Bose transmutation.
The {\em N-O} vortices are solitons
of the Maxwell-Higgs theory or the scalar electrodynamics
($SQED$), the relativistic generalization of Ginzburg-Landau
theory of superconductivity,
analogous to the Abrikosov vortices in a
type II superconductor.
It turns out that the path integral of the
Maxwell-Higgs system
can be expressed in terms of the {\em N-O} string
variables \cite{acpz}.
In this string representation a nontrivial long-range topological
interaction of {\em N-O} strings with particles
of electric charge $Qe$ appears
whenever $Qe$ is not an
integer multiple of $Ne$, the charge of the Higgs field.
This is a four-dimensional
analogue \cite{aw} of the Aharonov-Bhom effect.
This long-range interaction
involves the four-dimensional
analogue of the Gauss linking number for loops in
three dimensions
i.e. the linking number $\ell k(\sigma,j_C)$ of world
sheets of the strings
and the current $j_C$ of the charge $Qe$.
The effect of this linking number contribution
is to turn the bosonic propagator into a spinorial propagator.

	In this letter we consider the extension of
the previous results to the case
of the $compact$ version of the Maxwell-Higgs field theory
on the lattice, the $SQED_c$. Unlike the ordinary
(non-compact) $SQED$, the
path integral of this model can be
expressed in terms of {\em N-O} string vortices with
magnetic monopoles at their ends (see below).
One motivation to consider this extension
relies on the fact that this model contains the main
ingredients to reflect somehow the physics
of a real system.
We have in mind
a `C-shaped' ferromagnet with a slab
of a type-II superconductor inbetween.
In fact, quantum-mechanically,
a charge always sees `solenoids' irrespective
of the nature of magnetic sources \cite{ms}.
In other words, a charge would take a magnet for an
equivalent collection of thin solenoids or Dirac strings.
Then, provided a pair of monopole-antimonopole
embodied in the $SQED_c$ lattice model are fixed, it is
reasonable to regard them
as the poles of the ferromagnet.
The magnetic flux get trapped inside the superconductor
within the {\em N-O} strings which connect the magnetic poles
which are the equivalent of the the Abrikosov magnetic
flux tubes of the type II superconductor.
The external charge $Qe$
can be regarded as a fixed doping.

	The $SQED_c$ describes the interaction
of a compact gauge field $\theta_\mu \in (-\pi,\pi]$
with a scalar field $\phi=|\phi|e^{i\varphi}$
with charge $Ne$. The self-interaction of the
scalar field is given by the potential $V=\lambda
(|\phi|^2-a^2)^2$. For simplicitly, we
consider the limit $\lambda \rightarrow \infty$
which freezes the radial degree of freedom of the
Higgs field. This is not a strong
restriction, in fact it is known that
the numerical results already  obtained at
$\lambda=1$ are indistinguishable from the frozen case
\cite{jersak}. Thus the
dynamical variable is compact:
$\varphi \in (-\pi ,\pi]$.

Theories with Abelian compact variables contain topological
excitations: vortices in the $XY$ model,
monopoles in the pure compact electrodynamics,
{\em N-O} strings in the non-compact
scalar electrodynamics,
etc. There is one type of topological object for
each compact variable.
The $SQED_c$ model we are considering has two
compact variables
and therefore two types of topological
excitations: $N$-$O$ vortices and magnetic monopoles.
This model is known to possess three phases namely, confining,
Coulomb and Higgs \cite{jersak}. The Higgs phase
is the only that supports
both magnetic vortices and monopoles.

        There are several examples in the lattice field theory
such that a change of variables in the partition function allows for
a formulation of the theory in terms of the physical excitations.
Banks, Kogut and Myerson \cite{bkm} introduced a general tranformation
which gives rise to a description of any abelian gauge theory
with compact variables
in terms of variables on the dual lattice
associated with its corresponding topological excitations.
This was a generalization of the
technique used by Jose, Kadanoff, Kirkpatrick and Nelson \cite{jkkn}
to derive the Kosterlitz-Thouless \cite{kt}
phase transition in terms of vortices for the $D=2$ $XY$ model.
Given a $D$ dimensional lattice theory
with abelian compact variables on $c_{k-1}$ cells ($k=1$: sites
and spin theory, $k=2$: links and gauge theory, $k>2$ hypergauge theory),
by means of the $Banks-Kogut-Myerson$ transformation,
one arrives to the $topological$ expression of the partition
function given by
\begin{equation}
Z_T \propto \sum_{^*t_i} \prod_i
\exp
[-2\pi^2\alpha_i \sum_{c_{D-k-1}}\,^*t_i \hat{\Delta_i}\,^*t_i \, ],
\label{eq:topo}
\end{equation}
where the index $i$ number the topological excitations,
$\alpha_i$ and $\hat{\Delta}$ are their respective coupling
constant and propagator and
$\,^*t_i$ denotes an integer variable attached to the
$c_{D-k-1}$ cells of the dual lattice
which correspond to the
world "trajectories" of the $D-k-2$ dimensional topological excitations.
In four dimensions, one has world sheets for $k=1$ and world
trajectories for $k=2$.

        In order to express the path integral of
$SQED_c$ in terms of its topological excitations we
consider the partition function for the Villain \cite{v}
form of the lattice action \footnote{ We choose the
Villain form instead of the ordinary Wilson form only
for simplicity, with the Wilson action it is possible
to repeat all that we do here.} which is given by
\begin{eqnarray}
Z = \int_{- \pi}^{\pi} (D\theta ) \int_\pi^\pi (D\varphi ) \sum_{ n }
\sum_{l} \nonumber\\
 \exp
[-\sum_{c_2} \frac{\beta}{2} (\, d\theta +2\pi n \,)^2
-\sum_{c_1} \frac{\kappa}{2} (\, d \varphi + 2\pi l + N\theta \, )^2 ],
\label{eq:Villain}
\end{eqnarray}
where $D\theta$ ($D\varphi$) denotes the integral over all link $c_1$
(site $c_0$) variables $\theta$ (\,$\varphi$\,),
$\beta=\frac{1}{e^2}$ is the gauge
coupling constant, $\kappa$ is the Higgs coupling constant and
$n_{c_2}$ ($l_{c_1}$) are integer variables defined at the lattice
plaquettes $c_2$ (links $c_1$).
Fixing the gauge $d \varphi = 0$ and
parametrizing the $n$ variables as
\begin{equation}
n=n[m] + dq \, ,
\label{eq:n}
\end{equation}
where $n[m]$ is a solution of
\begin{equation}
dn[m] = m \, ,
\label{eq:dn}
\end{equation}
we can extend the compact variable $\theta \in (-\pi, \pi)$ to
a non-compact $A=\theta + 2\pi q$ $\in (-\infty,+\infty)$.
Performing the Gaussian integration and some algebra we get:
\begin{equation}
Z = \sum_{n} \sum_{l}
 \exp
[-2 \pi^2 \beta (\, d n, \frac{1}{\Box + M^2} d n \, )
-2 \pi^2 \kappa (\, Nn +dl, \frac{1}{\Box + M^2}\, Nn+dl \,),
\label{eq:Vinterm}
\end{equation}
where $M^2=N\frac{\kappa}{\beta}$ is the mass acquired by the
gauge field due to the Higgs mechanism, and
$(..,..) \, = \, \sum_{c_k}$.
{}From (\ref{eq:n}) and (\ref{eq:dn}):
\begin{eqnarray}
m=dn \, , \label{eq:mn}\\
\mbox{and taking} \nonumber \\
\sigma=Nn+dl,
\label{eq:sigma}
\end{eqnarray}
by performing a duality
transformation we arrive to the $topological$ representation
of the path integral
\begin{eqnarray}
Z_T = \sum_{\,^* m} \sum_{\,^* \sigma}
 \exp [\; -2 \pi^2 \beta (\,^*m \, , \frac{1}{\Box + M^2} \,^*m \, )  \nonumber
 \\
-2 \pi^2 \kappa (\,^*\sigma \, , \frac{1}{\Box + M^2}\, \,^*\sigma \, ) \; ],
\label{eq:Ztopo}
\end{eqnarray}
where the $\,^*$ denotes the Hodge dual,
the $\,^*m$ are integer closed 1-forms ($\partial^*m=\partial \partial n
\equiv 0$) on the dual lattice, which correspond
to magnetic monopole loops, and the $\,^*\sigma$ are integer
two-forms on the dual lattice whose border are the
monopole loops i.e. from (\ref{eq:mn}) and (\ref{eq:sigma}) we have
\begin{eqnarray}
\partial^* \sigma= \partial
(N^*n + \partial^* l) = N\partial^*n  \label{eq:border}\\
= N^* m \,.
\label{eq:mono}
\end{eqnarray}
Let us focus our attention for a moment on
the meaning of equation (\ref{eq:border}).
For example, in a type II superconductor $N=2$ (the charge of the
condensate of Cooper pairs is 2$e$) then the magnetic flux is
quantized as $\Phi_0=\frac{2\pi}{2e}$
\footnote{The flux in high $T_c$ superconductors
like the YBa$_2$Cu$_3$O$_7$ is also quantized in units
of the magnetic flux quantum
$\Phi_0=\frac{2\pi}{2e}$ \cite{gbdkmsw}. }
therefore the orbital
angular momentum of an electron around a unit vortex is
$l_z=\frac{1}{2}-e\frac{\Phi_0}{2\pi}=0$ and the composite
is a boson \cite{w} . On the other hand, the magnetic flux
carried by a Dirac string is an integer multiple of
$\frac{2\pi}{e}$. So, equation (\ref{eq:border})
shows that the magnetic field of a pair
monopole-antimonopole ( the ends of a Dirac string )
is squeezed in $two$ {\em N-O} string vortices.

        Thus, we have arrived to an
expression of the path integral
as a sum over surfaces $\,^*\sigma$ on the dual lattice
which are the world sheets of
string vortices (closed and open with magnetic
monopoles at their ends).
These strings are obtained by intersecting the
closed world sheets with a plane $t= constant$.
The quantum average of the Wilson loop for
the charge $Qe$, $W_Q(C)=\exp [\, iQ(A,j_C) \,]$,
can be written in the topological representation
, repeating the same steps which
lead from eq.(\ref{eq:Villain}) to eq.(\ref{eq:Ztopo}), as
\begin{eqnarray}
<W_Q(C)>_T
= \frac{1}{Z_T}
\sum_{\,^* m} \sum_{\,^* \sigma}
 \exp [\; -2 \pi^2 \beta (\,^*m \, ,
 \frac{1}{\Box + M^2} \,^*m \,)    \nonumber  \\
-2 \pi^2 \kappa (\,^*\sigma \, ,
\frac{1}{\Box + M^2}\, \,^*\sigma \,) \nonumber \\
- \frac{Q^2e^2}{2} (\, j_C \, ,
\frac{1}{\Box+M^{2}} j_C \,) \nonumber \\
+2\pi i \frac{Q}{N} (\, j_C \, , \frac{1}{\Box+M^2}
\partial \sigma \, )\; ].
\label{eq:WJ}
\end{eqnarray}
Using that $\partial j_C$ = 0 we have that
\begin{equation}
(\, j_C \, , \frac{1}{\Box+M^2} \partial \sigma \, )
= (\, j_C \, ,  l \,) - M^2 (\,j_C \, , {\Box}^{-1} l \,),
\end{equation}
and then we get

\newpage

\begin{eqnarray}
<W_Q(C)>_T=
\sum_{\,^* n} \sum_{\,^* l}
 \exp [\; -2 \pi^2 \beta (\,^*m \, ,
 \frac{1}{\Box + M^2} \,^*m \,)    \nonumber  \\
-2 \pi^2 \kappa (\,^*\sigma \, ,
\frac{1}{\Box + M^2}\, \,^*\sigma \,) \nonumber \\
- \frac{Q^2e^2}{2} (\, j_C \, ,
\frac{1}{\Box+M^{2}} j_C \,) \nonumber \\
-2\pi i Q\frac{\kappa}{\beta} (\, j_C \, ,
\frac{1}{\Box+M^2} l \, ) \nonumber \\
+2\pi i \frac{Q}{N} \ell k(\sigma_0 \, ,j_C) \;] \, .
\label{eq:WJ2}
\end{eqnarray}
The first four terms in the exponent describe short range (Yukawa)
interactions. The last long-range term
is the linking number $\ell k(\sigma_0,j_C)$ of the
the current $j_C$ generated by the charge $Qe$
and the world-sheets $^*\sigma_0=\,^*\sigma-N^*n$ which
are closed surfaces formed by two open pices: the world-sheets
of the {\em N-O} and the world-sheets of the Dirac
strings. The explicit expression of $\ell k$ is
\begin{equation}
\ell k = (\;^*j_C,^*l \;)=(\;^*j_C , \frac{1}{\Box} d\,^*(\sigma-Nn) \;).
\label{eq:linking}
\end{equation}

        Therefore,  if we neglect the Yukawa short-range
interaction terms of the Wilson loop average (\ref{eq:WJ2})
we get an expression that has exactly the form
obtained in Ref.\cite{fgt} from the (3+1) generalization of
the (2+1) Polyakov's \cite{p} construction which included a
bosonic-string in electromagnetic interaction described
by a potential $A_\mu$ and
under the action of a Kalb-Ramond
antisymmetric background field $B_{\mu \nu}$.
These fields are coupled through a four-dimensional
analogous to the Chern-Simons interaction:
\begin{equation}
\int  B \wedge F,
\end{equation}
where $F$ is the two-form electromagnetic
field strength.
The path integral (in the continuoum)
of this system may be cast in the form:
\begin{equation}
Z = \sum_{\sigma} e^{-TS(\sigma)}\exp
[ -\frac{i}{4\pi^2} \oint_C
dx\int_{\sigma_C}d\sigma_C(x')^*
( d\frac{1}{|x-x'|^2})\,].
\label{eq:cont}
\end{equation}
As in Ref.\cite{p}, the integral in the exponential is a
topological number: the self-linking number
$\ell k(\sigma_C,C)$ which
measures the number of times the closed path $C$ links the
sheet $\sigma_C$ with border $C$ in four dimensions.
Thus there is a difference
between the expression we have obtained here,
involving a linking number, and that
of (\ref{eq:cont}).
In fact, the four-dimensional extension of
Polyakov's construction leads to a
singular expression of the self-linking number,
and some regularization procedure or framing is needed.
The regularized self-linking number
is equal to the sum of the twist of the framed loops plus the
writhe of the original loop.
Polyakov neglects in the regularization procedure
the contribution of the writhe and he only considers the twist
contribution of the linking number. A ratianale for this choice
is that the writhe contribution is related with orbital effects
and not with intrinsic angular momentum.
Here, we proceed with the same philosophy.
An analogous situation seems to arise in
when one studies the transmutation phenomena
following the Wilczek approach \cite{w2} and consider
the `return' flux which closes the vortex line \cite{pe}.
Then, one has two vortices with opposite magnetic field and
both contribution to the angular momentum of the composite
cancel out.

        Now, one can  prove the transmutation,
whenever $\frac{Q}{N} \neq \;integer$,  by
following the same steps of \cite{fgt}. The main idea is that the
action appearing in (\ref{eq:cont}) leads after a Dirac quantization to
a set of variables that behave as Pauli matrices and reproduce the
propagator of the fermionic string.


One could have followed a similar approach in the (2+1) case.
Starting from the charge vortex pair, one could have studied the
transition amplitude by computing the Feynman path integral of a charged
particle in the presence of a magnetic vortex. This computation would
have led to the ordinary Gauss linking number and to the action already
considered in Ref. \cite{p}.

Sumarizing, we have proved that the open string-like
excitations of the
Maxwell-Higgs system in presence of an external
charge undergo Fermi-Bose transmutation.
The possibility to connect this result with a real system
relies on the fact that, from the quantum-mechanical
point of view, a charge interacts with a magnetic source
only through the vector potential. Thus,
a charge always sees a magnet
as an equivalent collection of thin solenoids
or Dirac strings. The fact that in this case the
topological contribution arises from a linking number
instead of a self-linking number seems to be
irrelevant due to the fact that
the vortices in a type II superconductor, even in
the high $T_c$ superconductors \cite{yao},
only have small displacements and therefore
one can consider that the world-line of the (fixed) charge
remains close to the vortex world-sheet.
However,
it is still not completely clear if the transmutation here
observed could affect the vortex dynamics and
the behaviour of a superconductor.


\newpage

\begin{thebibliography}{99}

\bibitem{kr} M. Kalb and P. Ramond, {\em Phys. Rev.}
                {\bf D9} (1974) 2273.

\bibitem{gs} R. Gambini and L. Setaro, {\em Phys. Rev. Lett.}
{\bf 65} (1990) 2623.

\bibitem{fg} H. Fort and R. Gambini, Preprint IFFC-95-01,
             Feb. 1995, hep-th/9502008.

\bibitem{no}   H.B.Nielsen and P.Olesen, {\em Nucl. Phys.}
                      {\bf B61} (1973) 45.

\bibitem{acpz} E. T. Akhmedov, M. N. Chernodub, M.I.Polikarpov,
and M.A.Zubkov, Preprint ITEP-95-24, June 1995, hep-th/9505070.

\bibitem{aw} M. G. Alford and F. Wilczek, {\em Phys. Rev. Lett.}
{\bf 62} (1989) 1071.

\bibitem{ms} J. Maeda and K. Shizuya, Z. {\em Phys.} {\bf C60}
                     (1993) 265.


\bibitem{jersak} K. Jansen, J. Jersak, C.B. Lang, T. Neuhaus
and G. Vones, {\em Nucl. Phys.} {\bf B265} (1986) 129.

\bibitem{bkm}  T. Banks, R. Myerson and J.B. Kogut,
                {\em Nucl. Phys.} {\bf B129} (1977) 493.

\bibitem{jkkn}  J. Jose, L-P. Kadanoff, S. Kirkpatrick and D.R.
                Nelson, {\em Phys. Rev.}{\bf B16} (1977) 1217.

\bibitem{kt}   J. Kosterlitz and D. Thouless, {\em J.Phys.}
                {\bf C6} (1973) 1181.

\bibitem{v}    J. Villain, {\em J. Phys.} (Paris)
               {\bf 36} (1975) 581.

\bibitem{gbdkmsw} P. L. Gammel, D. J. Bishop, G. J. Dolan,
               J. R. Kwo, C. A. Murray, L. F. Schneemeter and
               J. V. Waszczak, {\em Phys. Rev. Lett.} {\bf 59}
               (1987) 2592.

\bibitem{w} F. Wilczek, {\em Phys. Rev. Lett.}
            {\bf 48} (1982) 1144.

\bibitem{fgt} X. Fustero, R. Gambini and A. Trias,
              {\em Phys. Rev. Lett.} {\bf 62} (1989) 1964.

\bibitem{p} A. M. Polyakov, {\em Mod. Phys. Lett.}
            {\bf A3} (1988) 325.


\bibitem{w2} F. Wilczek, {\em Phys. Rev. Lett.}
             {\bf 49} (1982) 957.

\bibitem{pe} M. Peshkin, {\em Phys. Rep.} {\bf 80} (1989) 376.

\bibitem{yao} Z. Yao, S. Yoon, H. Dai, S. Fan and C. M. Lieber,
{\em Nature} {\bf 371} (1994) 777.

\end{thebibliography}
\end{document}